\documentclass[a4paper,UKenglish,cleveref, autoref, thm-restate]{lipics-v2021}

\usepackage{booktabs}

\title{Distributed Incremental {SAT} Solving with {Mallob}:\\
	Report and Case Study with Hierarchical Planning} 

\titlerunning{Distributed Incremental SAT Solving with Mallob} 

\author{Dominik Schreiber}{Karlsruhe Institute of Technology, Karlsruhe, Germany \and \url{https://satres.kikit.kit.edu/schreiber/} }{dominik.schreiber@kit.edu}{https://orcid.org/0000-0002-4185-1851}{}

\authorrunning{D. Schreiber} 

\Copyright{D. Schreiber} 



\category{} 

\relatedversion{} 

\nolinenumbers 

\hideLIPIcs  

\EventEditors{John Q. Open and Joan R. Access}
\EventNoEds{2}
\EventLongTitle{42nd Conference on Very Important Topics (CVIT 2016)}
\EventShortTitle{CVIT 2016}
\EventAcronym{CVIT}
\EventYear{2016}
\EventDate{December 24--27, 2016}
\EventLocation{Little Whinging, United Kingdom}
\EventLogo{}
\SeriesVolume{42}
\ArticleNo{23}

\newcommand{\systemname}[1]{\mbox{\textsc{#1}}}

\newcommand{\glucose}{\systemname{Glucose}}

\newcommand{\lilotane}{\systemname{Lilotane}}
\newcommand{\lingeling}{\systemname{Lingeling}}

\newcommand{\mallob}{\systemname{Mallob}}
\newcommand{\mallobsat}{\systemname{MallobSat}}

\newcommand{\cadical}{\systemname{CaDiCaL}}

\newcommand{\mallotane}{\systemname{Mallotane}}

\newcommand{\whp}{w.h.p\@ifnextchar.{\@}{.\@\xspace}}
\newcommand{\iid}{i.i.d\@ifnextchar.{\@}{.\@\xspace}}
\newcommand{\eg}{eg\@ifnextchar,{\xspace}{,\xspace}}
\newcommand{\ie}{ie\@ifnextchar,{\xspace}{,\xspace}}
\newcommand{\cf}{\emph{cf.}} 

\newcommand{\pyplotscale}{0.9}

\newcommand{\compclass}[1]{{\sffamily #1}}

\DeclareTextFontCommand{\hyphenedtexttt}{\ttfamily\hyphenchar\font=45\relax}
\newcommand{\plandomain}[1]{\hyphenedtexttt{#1}}

\begin{document}

\maketitle

\begin{abstract}
This report describes an extension of the distributed job scheduling and SAT solving platform \mallob{} by \emph{incremental SAT solving}, embedded in a case study on SAT-based hierarchical planning.
We introduce a low-latency interface for incremental jobs and specifically for {IPASIR}-style incremental {SAT} solving to \mallob{}.
This also allows 
to process many independent planning instances in parallel via \mallob{}'s scheduling capabilities.
In an experiment where 587 planning inputs are resolved in parallel on 2348 cores, 
we observe significant speedups for several planning domains where {SAT} solving constitutes a major part of the planner's running time.
These findings indicate that our approach to distributed incremental SAT solving may be useful for a wide range of {SAT} applications.
\end{abstract}

\section{Introduction}

Propositional satisfiability (SAT) solving is a crucial automated reasoning tool~\cite{biere2009handbook}.
In prior works, we explored how distributed environments (like HPC and clouds) can be exploited effectively for SAT solving---both with (massively) parallel SAT solving and by processing many tasks in parallel---which has resulted in the distributed job scheduling and SAT solving platform \mallob{}~\cite{sanders2022decentralized,schreiber2021scalable}.
We also explored \textit{Totally Ordered Hierarchical Task Network} (TOHTN) planning and devised a novel SAT-based TOHTN planning approach, \lilotane{}, which omits a central stage of prior SAT-based approaches~\cite{schreiber2021lilotane}.
\lilotane{} crucially builds upon \emph{incremental SAT solving}, i.e., interactive sequences of SAT queries on successively expanding formulas~\cite{balyo2016sat,een2003temporal}.
However, there has been little work on parallel incremental SAT solving~\cite{wieringa2013asynchronous} and (to our knowledge) none on distributed incremental SAT solving.

This report presents a study in which we extend \mallob{} by full featured {incremental SAT solving}, using the previously developed hierarchical planning approach as an example application.
As such, we not only explore a possible avenue to parallelizing TOHTN planning---a challenging problem due to its \compclass{2-EXPTIME}-hard complexity~\cite{alford2015tighthtn}---but we also enhance and evaluate the practical merit of our distributed platform from the perspective of an application of incremental SAT solving and thus provide an outlook on how such applications may be able to profit from \mallob{} in the future.

We first analyze the data gathered in earlier evaluations of \lilotane{}~\cite{schreiber2021lilotane} and investigate requirements for an effective parallelization of \lilotane{}'s SAT solving backend.
We conclude that an incremental job interface with very low latencies is needed to account for the many trivial SAT calls made by applications like \lilotane{}.
We then enhance \mallob{} by such a low-latency interface for incremental jobs and extend \mallobsat{} to support incremental SAT solving.
We conduct an experiment where we use a single large-scale run of \mallob{} to process many planning tasks at the same time.
We observe noticeable speedups on the inputs on which the planning approach makes sufficiently difficult SAT calls.

\textbf{Context.} This report is a lightly edited standalone version of the (otherwise unpublished) Chapter~7 of the author's dissertation~\cite{schreiber2023scalable}
and is meant primarily as a reference and data-backed documentation of \mallob{}'s incremental SAT solving capabilities.
This work was authored in 2023 and thus does not discuss any later related works.
A preliminary form of the presented parallelization of TOHTN was featured before in a student research project~\cite{bretl2022parallel}. 


We assume the reader to be familiar with (distributed, incremental) SAT solving in general and with the \mallob{} system in particular and refer to according publications for the relevant preliminaries and background~\cite{sanders2022decentralized,schreiber2021scalable}.

\section{Requirement Analysis}  \label{lilotane-mallob:requirement-analysis}

In the following, we analyze the data gathered in earlier evaluations of \lilotane{} with respect to the opportunities and challenges of parallelizing its SAT solving calls.

Each SAT solving call in a \lilotane{} planning procedure corresponds to one so-called \textit{hierarchical layer} of the input (see ref.~\cite{schreiber2021lilotane} for details).
As such, \lilotane{} emits a series of unsatisfiable formulas until it reaches a layer where a plan can be found and the encoding is satisfiable.
Since different domains result in task networks of varying depth, this number of calls varies significantly across domains.
For the median instance of the 2020 International Planning Competition (IPC)~\cite{behnke2020international}, \lilotane{} makes ten SAT solving calls.
Instances from the \plandomain{Childsnack} domain, which feature a shallow and fixed hierarchy, always result in exactly three calls.
The other extreme is the \plandomain{Towers} domain at 120.6 calls on average.

\begin{figure}[b!]
	\begin{minipage}{0.5\textwidth}
		\centering
		\includegraphics[scale=\pyplotscale]{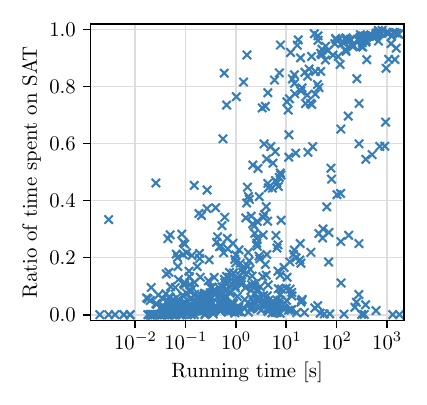}	
	\end{minipage}%
	\begin{minipage}{0.5\textwidth}
		\centering
		\includegraphics[scale=\pyplotscale]{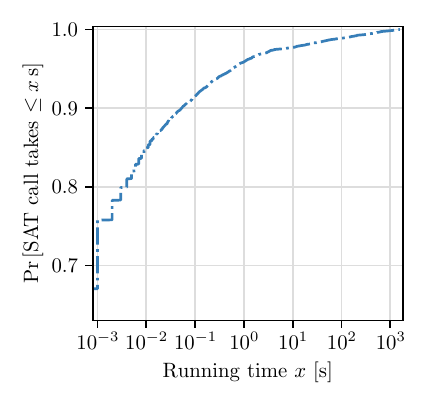}	
	\end{minipage}
	\caption{Analysis of \lilotane{} on IPC 2020 instances (based on data from ref.~\cite{schreiber2021lilotane}). 
		Left: Ratio of time spent on SAT solving for solved instances.
		Right: CDF for the duration of individual SAT calls on solved instances (note the offset of the $y$ axis).
		Both $x$ axes are scaled logarithmically.}
	\label{fig:lilotane-scatter-sat-ratio}
\end{figure}

In terms of the \textit{total} time which \lilotane{} spent on successfully solved instances from the IPC 2020 benchmarks, around 78\% was spent on SAT solving itself. 
On the other hand, considering the \textit{median} instance, only 9.5\% of time was spent on SAT solving.
Fig.~\ref{fig:lilotane-scatter-sat-ratio} (left) shows this ratio for every IPC instance solved by \lilotane{}.
We found a weak to moderate correlation between an instance's running time and the ratio of time spent on SAT solving (Pearson's $r=0.497$).
Such a correlation is desirable since difficult instances with long running times bear the highest potential for meaningful speedups. 
On 75\% of instances which take at least 20\,s to solve, \lilotane{} spends the majority of time on SAT solving.
The same holds for 80\% of the instances which take at least 100\,s to solve.
From a domain-dependent view, only two domains (\plandomain{Logistics} and \plandomain{Assembly}) feature such a SAT solving ratio larger than 50\% for the median instance.
Eight domains do not feature a single instance where this ratio even exceeds 10\%.
As such, the effectiveness of a parallelization of \lilotane{}'s SAT backend will be highly domain-dependent.
For fully scalable TOHTN planning across the entire range of available domains, \lilotane{}'s instantiation and encoding approach would need to be parallelized as well, which we do not consider in this work.


Another point to consider is that the time \lilotane{} spends on SAT solving is split into thousands of individual SAT calls, many of which are made in rapid succession and only few of which actually prove to be non-trivial.
Fig.~\ref{fig:lilotane-scatter-sat-ratio} (right) shows that two thirds of all SAT calls take less than a millisecond and more than 95\% of SAT calls take less than a second.
Therefore, in order to achieve good average speedups with a parallel SAT solving backend, it is crucial to carefully engineer an efficient interface with low turnaround times for individual SAT calls.
Parallelization should be performed in a \textit{cautious} manner---only investing work in scaling up a SAT call if it proves to be non-trivial.
The job model of \mallob{} is specifically designed for these criteria, offering low-latency scheduling of incoming jobs and a flexible amount of parallelization based on the system state and a job's perceived difficulty.

\section{Approach} \label{lilotane-mallob:approach}

In the following, we describe how we extended \mallob{} to support incremental (SAT) jobs and how we designed the interface between a \mallob{} process and an application.

\subsection{Incremental Jobs in \mallob{}}

Applications of incremental SAT, such as \lilotane{}, expect to interact with an incremental SAT solver.
It is possible to simulate an incremental SAT solving interface with a non-incremental solver by adding assumptions as permanent unit clauses and by in turn starting a completely new solver instance for each increment~\cite{nadel2012efficient}.
However, our requirement analysis suggests that 
our application relies on very low response times, which likely renders repeated re-initialization of solver threads prohibitively costly.
For this reason and also to allow many further applications of incremental SAT solving~\cite{axelsson2008analyzing,gocht2017accelerating,buning2019using,liu2016oracle,surynek2022migrating} to use our system efficiencly, we introduce a ``native'' pipeline for incremental job processing to \mallob{}.

We model a simple protocol for incremental jobs in \mallob{} as follows:
A job can have multiple \emph{revisions} (which correspond to individual incremental SAT solving calls) indexed by $r=0,1,\ldots$.
Each revision $r$ adds a certain payload $P_r$ to the job and can be submitted as soon as the previous revision $r-1$ has finished (or was cancelled).

Recall that \mallob{} follows a \emph{malleable job scheduling} model where the processes belonging to a job are organized as a \textit{binary tree of workers} that may grow, shrink, and re-grow differently throughout the computation~\cite{sanders2022decentralized}.
The worker at the root of an incremental job's tree is present constantly, i.e., from the initiation of revision 0 until the conclusion of the final revision (which is indicated by a special signal from the user).
Whenever some revision $r-1$ is done, the tree shrinks to a single worker, i.e., all non-root workers leave the tree structure and are suspended.
The tree re-grows when revision $r$ arrives.
\mallob{}'s according protocols preferably reactivate suspended workers and thus re-use initialized resources from earlier (e.g., solver threads, payloads of past revisions) rather than creating new workers at different processes~\cite{sanders2022decentralized}.
Revision payloads are transferred to workers down the job tree just like usual job descriptions, with the constraint that a worker can only begin to process revision $r$ after all payloads $P_0, \ldots, P_r$ have arrived at this worker.

\subsection{Interface}

In the scope of this study, we devised an application interface for incremental SAT solving with \mallob{} that is fully generic, i.e., that can be linked with any application of plain incremental SAT solving without any notable code modifications.
Note that even better performance may be achievable by compiling application code directly into \mallob{}'s client code and thus avoiding the overhead introduced by inter-process communication.

\begin{figure}[h]
	\centering
	\includegraphics[width=\textwidth]{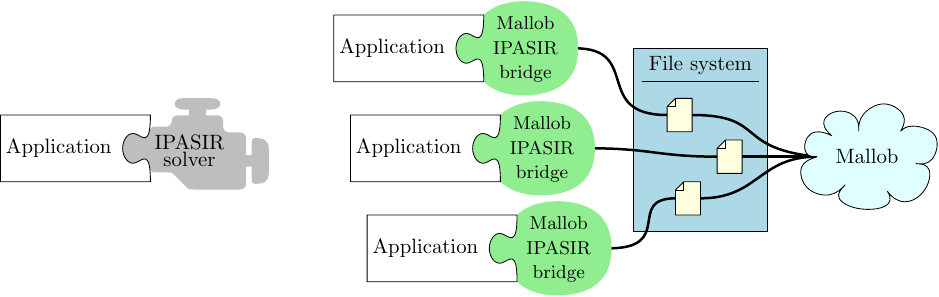}
	\caption{Incremental SAT application interface.
		Left: conventional IPASIR interface with direct linkage of the application code to SAT solver code.
		Right: our approach.
		The application is linked with bridge code, which communicates with a running instance of \mallob{} via the local machine's file system and via inter-process communication.}
	\label{fig:mallob-ipasir}
\end{figure}

Fig.~\ref{fig:mallob-ipasir} illustrates our technical setup for incremental SAT solving.
We emulate the incremental SAT solver interface IPASIR~\cite{balyo2016sat} for the application side and thus provide a drop-in replacement for sequential IPASIR solvers.
On a technical level, an application process communicates with \mallob{} by writing a small JSON file with the job's meta data to a directory monitored by a local \mallob{} process.
The job description itself is then transferred from the application to the \mallob{} process via inter-process communication.
The application then waits for the \mallob{} process to transfer back a result.
We use named (UNIX) pipes for inter-process communication to ensure fast transfer of large amounts of data (i.e., formulas and satisfying assignments).
We use the Linux subsystem \texttt{inotify} for watching files in a directory, which allows \mallob{} to react to incoming jobs immediately.

\subsection{Incremental SAT Solving Engine}

In order to prepare \mallobsat{}~\cite{schreiber2021scalable} for incremental SAT solving, we restrict our portfolio to \cadical{}~\cite{biere2018cadical} and \lingeling{}~\cite{biere2016splatz}, which both natively support incremental SAT solving.
We perform clause sharing just like for non-incremental solving.
The only SAT-specific invariant we need to maintain is that a solver never imports any clauses from a future revision $r' > r$ with respect to its own current revision $r$.
We achieve this by marking each contribution to a clause buffer with the current revision of the exporting SAT process.
When aggregating buffers, the maximum revision that is present is propagated and is then broadcast together with the final buffer.


\subsection{Reducing Revision Turnaround Times}

Our analyses of \lilotane{}'s behavior on IPC instances showed that low latencies of individual SAT solving calls are crucial for performance.
We made a few modifications to \mallob{} to account for this requirement and to minimize the turnaround time of a trivial SAT call.

First, we remove large parts of the initial scheduling latency of a job for subsequent revisions by preserving the root worker of each job across revisions.
To introduce a new revision for a job, only a single message (featuring the new revision payload) must be sent to this root worker.
Additional workers are then allocated as usual by emitting a balancing event, computing new fair job volumes, and emitting job request messages (see~\cite{sanders2022decentralized}).

Secondly, \mallob{}'s main thread usually checks the status of a job every 10\,ms.
This interval is now set to 1\,ms whenever a new worker is scheduled on the process and then increases successively until it reaches 10\,ms.
Thirdly, we reduced the latency of a SAT solving process reporting a trivial result by letting \mallob{}'s main thread read small job results itself immediately instead of always running a separate background worker for this task.

\section{Evaluation}  \label{lilotane-mallob:evaluation}

We now evaluate our approach.
Our software and experimental data are available online.\footnote{\url{https://zenodo.org/doi/10.5281/zenodo.10184679}}

All experiments have been conducted on the supercomputer \textbf{SuperMUC-NG}.
This system at the \textit{Leibniz Rechenzentrum} features a total of ``\textit{311\,040 compute cores with a main memory of 719\,TB and a peak performance of 26.9\,PetaFlop/s}''~\cite{lrz2023supermucng}.
In particular, SuperMUC-NG features 6\,336 ``thin'' compute nodes each with a two-socket Intel Skylake
Xeon Platinum 8174 processor clocked at 2.7$\,$GHz with 48 physical cores (96 hardware threads) and 96\,GB of main memory.
Nodes are interconnected via OmniPath~\cite{birrittella2015intel}. 

As a sequential baseline, we run \lilotane{} on the IPC 2020 benchmarks.
We changed \lilotane{}'s backend from \glucose{} to \cadical{} for improved performance~\cite{supervisedthesis:schnell}.

We refer to our approach as \mallotane{}~\cite{bretl2022parallel}---a portmanteau of \mallob{} and \lilotane{}.
We use a subset of the IPC 2020 benchmarks, filtering out two kinds of instances based on our sequential baseline: (a) instances where \lilotane{} exceeded the main memory limit of 8\,GB, and (b) instances where the time spent on SAT solving accounts for less than 0.1\% of the total running time.
34.2\% of all instances match these criteria, leaving 587 instances for our experiment.
Evidently, to improve performance on the filtered instances, more invasive changes to \lilotane{} itself would be in order.
Note that the remaining benchmark set still features instances from several domains where the amount of time spent on SAT solving is negligible, therefore allowing us to analyze the worst-case behavior of \mallotane{}.

We construct a scenario as in ref.~\cite{schreiber2021scalable}: We allocate a fixed amount of computational resources and attempt to solve as many problems as possible within a given time frame.
Since we only parallelize parts of each planner, we decided to set up our experiment in such a way that each planner instance initially receives very few resources (four cores) and then successively scales up to 30-40 cores given that the task is sufficiently difficult.
Specifically, we configure each job to grow by another job tree layer every 0.5 seconds: We initialize job demand $d_j$ (the desired number of processes~\cite{sanders2022decentralized}) to 1 at the arrival of $j$ and update $d_j:=2\cdot d_j+1$ every 0.5 seconds until $d_j$ has reached the number of processes in the system.

Deploying \lilotane{} and \mallob{} at the same time on SuperMUC-NG proved to be challenging due to the small amount of main memory available (2\,GB per core or 1\,GB per hardware thread).
Since one instance of \lilotane{} on average requires considerably more memory than one SAT solver instance in our system, we decided to deploy \lilotane{} and \mallob{} as follows:
\begin{itemize}
	\item We allocate $4 \times 587 = 2\,348$ cores of 49 compute nodes, spawning one \mallob{} process for each set of four cores.
	\item We configure \mallobsat{} to only employ \textit{three} solver threads per process.
	\item Together with each \mallob{} process we execute one instance of our planner, which then communicates with the \mallob{} process as a SAT interface.
	Note that this \mallob{} process only introduces the corresponding job to the decentralized scheduler---the job's first worker may be scheduled to \emph{any} \mallob{} process.
\end{itemize}
Note that, in order to avoid allocating additional compute nodes, we distributed the planning problems over the compute nodes in a balanced manner based on the amount of memory required by sequential \lilotane{}.
In a real-world application of our system, we would expect the application code (i.e., our planners) to be run on client machines that are separated from \mallob{}'s worker processes.


\subsection{Results}

\begin{table}[b!]
	\centering
	\footnotesize
	\addtolength{\tabcolsep}{-0.3em}
	\begin{tabular}{lrrrrrr}
		\toprule
		Domain & \# & min. & med. & geom. & max. & total \\
		\midrule
		Assembly       	& 5   & 0.15 & 1.11 & 0.772 & 3.19  & 2.821 \\
		Barman          & 18  & 0.18 & 0.84 & 0.808 & 4.12  & 2.069 \\
		BlocksworldG    & 22  & 0.12 & 0.81 & 0.639 & 1.14  & 0.946 \\
		BlocksworldH    & 1   & 0.91 & 0.91 & 0.910 & 0.91  & 0.914 \\
		Childsnack      & 26  & 0.19 & 0.46 & 0.460 & 1.09  & 0.930 \\
		Depots          & 22  & 0.32 & 0.77 & 0.734 & 1.33  & 1.196 \\
		Elevator    	& 146 & 0.14 & 3.40 & 3.103 & 24.82 & 9.054 \\ 
		Entertainment   & 2   & 0.93 & 0.99 & 0.960 & 0.99  & 0.938 \\
		Factories       & 4   & 0.16 & 1.75 & 0.839 & 2.06  & 1.835 \\
		Freecell   		& 12  & 1.01 & 2.28 & 2.429 & 5.17  & 3.228 \\
		Hiking          & 23  & 0.29 & 0.82 & 0.815 & 2.06  & 1.455 \\
		Logistics  		& 50  & 0.54 & 2.02 & 1.948 & 3.86  & 2.171 \\
		MinecraftP      & 4   & 0.78 & 2.16 & 1.717 & 2.67  & 2.294 \\
		MinecraftR      & 32  & 0.37 & 0.96 & 0.895 & 1.09  & 0.990 \\
		MonroeFO     	& 19  & 0.58 & 0.91 & 0.899 & 1.14  & 0.949 \\
		MonroePO 	    & 19  & 0.83 & 0.94 & 1.014 & 2.68  & 1.058 \\
		Multiarm        & 4   & 1.02 & 2.20 & 1.649 & 3.17  & 2.869 \\
		Robot           & 8   & 0.10 & 0.34 & 0.375 & 1.37  & 1.249 \\
		Rover           & 24  & 0.12 & 1.89 & 1.528 & 5.85  & 3.477 \\
		Satellite       & 15  & 0.16 & 1.04 & 0.849 & 1.42  & 1.125 \\
		Snake           & 20  & 0.14 & 0.74 & 0.586 & 1.11  & 1.027 \\
		Towers          & 6   & 0.20 & 0.84 & 0.581 & 0.89  & 0.864 \\
		Transport       & 32  & 0.13 & 0.54 & 0.645 & 8.63  & 3.207 \\
		Woodworking     & 22  & 0.14 & 0.65 & 0.547 & 1.01  & 0.824 \\
		\bottomrule
	\end{tabular}
	\caption{Speedups of \mallotane{} over \lilotane{} by domain on the commonly solved instances from the considered subset of IPC 2020 benchmarks, in terms of minimum, median, geometric mean, maximum, and total speedup.}
	\label{tab:mallotane-speedups}
\end{table}

\begin{figure}
	\centering
	\includegraphics[scale=\pyplotscale]{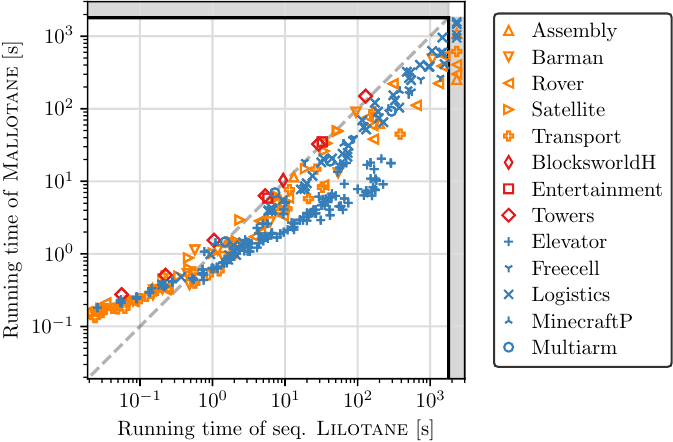}
	\caption{Selected (log-log) running times of \lilotane{} vs. \mallotane{},
		showing five domains with median speedup $\geq 2$ (blue), three domains where no instance was sped up (red), and five domains with notable outliers (orange).}
	\label{fig:mallotane-scatter}
\end{figure}

We now discuss the results of our experiment.
Tab.~\ref{tab:mallotane-speedups} lists the speedups of \mallotane{} over \lilotane{} and Fig.~\ref{fig:mallotane-scatter} shows a visual comparison of individual running times.
As expected, our setup incurs some overhead compared to a sequential SAT solver that is compiled into the planning application.
This overhead, however, proves to be small and mostly concerns easy problems which are resolved in less than a second.
Whether \mallotane{} results in speedups and how high these speedups can reach is, unsurprisingly, highly dependent on the planning domain.
We observed a median speedup greater than two for the five domains \plandomain{Elevator}, \plandomain{Freecell}, \plandomain{Logistics}, \plandomain{Minecraft-Player}, and \plandomain{Multiarm-Blocksworld}.
\mallotane{} achieved the highest domain-specific performance for \plandomain{Elevator} at a median speedup of 3.4 and also achieved the largest individual speedup (24.8) on a problem from this domain.
In addition, \mallotane{} is able to solve ten additional instances, specifically from the domains \plandomain{Logistics} (5), \plandomain{Assembly} (2), \plandomain{Rover} (2), and \plandomain{Transport} (1).
There are three domains on which \mallotane{} never achieves any speedups, namely \plandomain{BlocksworldH}, \plandomain{Entertainment}, and \plandomain{Towers} (which together only amount to nine considered instances).
For a total of 15 domains, the median speedup achieved is less than one, indicating a slowdown for the majority of instances from these domains.
\plandomain{MinecraftR} is one of them and constitutes a good example for a domain where very little time is required for SAT solving.
On eight instances in this domain, \lilotane{} spent less than 1\% of its time on SAT solving with running times ranging between 12 and 89 seconds.
\mallotane{}'s ``speedups'' range between 0.924 and 0.972 on these instances, indicating an overhead between 2.8\% and 7.6\%.

As Fig.~\ref{fig:mallotane-scatter} shows, almost all instances which result in a considerable slowdown are solved very quickly in terms of absolute running times.
As such, while the median speedup of \mallotane{} on the considered benchmark set is 1.018, its total speedup is 2.63.
\begin{figure}[t]
	\centering
	\includegraphics[scale=\pyplotscale]{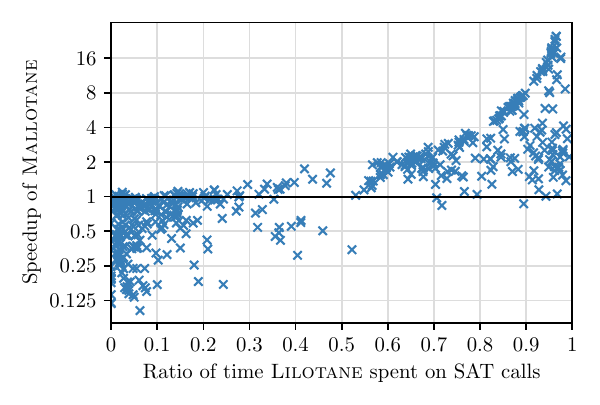}
	\caption{Speedups of \mallotane{} (\mbox{log.} scale) relative to the ratio of time sequential \lilotane{} spent on SAT calls, with $y=1$ highlighted by a black line.}
	\label{fig:mallotane-speedup-by-seqsatratio}
\end{figure}
In addition, we set each speedup in relation to the ratio of time (sequential) \lilotane{} spent on SAT calls.
Fig.~\ref{fig:mallotane-speedup-by-seqsatratio} confirms that most slowdowns occur on instances where SAT solving constitutes a small ratio of the overall running time.
Indeed, whether or not \mallotane{} achieves a speedup $>1$ can be predicted rather accurately based on whether sequential \lilotane{} spends the majority of its time on SAT solving.
Such a classifier wrongly attributes no speedup to 41 instances with some speedup and wrongly attributes a speedup to 4 instances with no speedup.
All remaining 491 instances are classified correctly.

\begin{figure}
	\centering
	\includegraphics[scale=\pyplotscale]{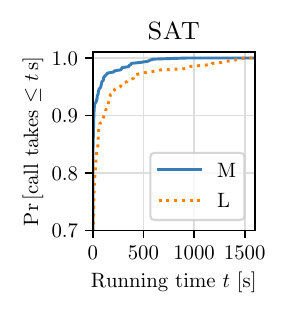}%
	\includegraphics[scale=\pyplotscale]{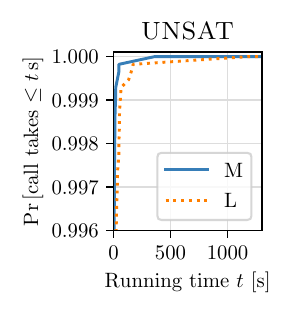}%
	\includegraphics[scale=\pyplotscale]{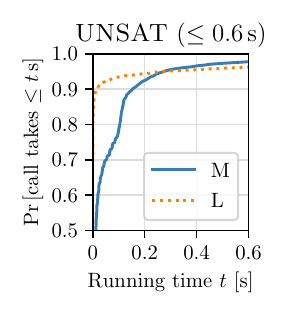}%
	\caption{Distribution over the duration of individual SAT solving calls for sequential \lilotane{} (L) and for \mallotane{} (M), separated into SAT and UNSAT results, for commonly solved instances only. The $y$ axes are offset to allow for discerning relevant differences. Since almost all UNSAT calls are very short, we added a separate plot only showing calls which take less than 0.6\,s.}
	\label{fig:mallotane-satcall-cdfs}
\end{figure}

On the 536 commonly solved instances, both the sequential and the parallel approach made a total of 6181 SAT solving calls.
5645 calls resulted in unsatisfiability (``UNSAT calls'') and the remaining 536 in satisfiability (``SAT calls'').
Fig.~\ref{fig:mallotane-satcall-cdfs} shows how the duration of these calls is distributed for the sequential and the parallel backend.
With the sequential backend, 96.86\% of UNSAT calls and 57.65\% of SAT calls took less than one second---confirming that the performance of our planner crucially depends on low overhead for individual calls.
The break-off point where our \mallob{} backend begins to outperform the sequential backend is at 0.23\,s:
The same ratio of calls (90.4\%) finished within this time for both approaches.
For UNSAT calls this break-off point is at 0.28\,s (95.1\% of calls finished) while for SAT calls it is already at 0.053\,s (31.7\% of calls finished).
Still, since both satisfiable and unsatisfiable increments can be very hard, the total speedup is comparable for SAT calls vs. UNSAT calls (3.257 vs. 2.908) and, notably, matches the three SAT solving threads per input.

\begin{figure}[t]
	\centering
	\includegraphics[scale=\pyplotscale]{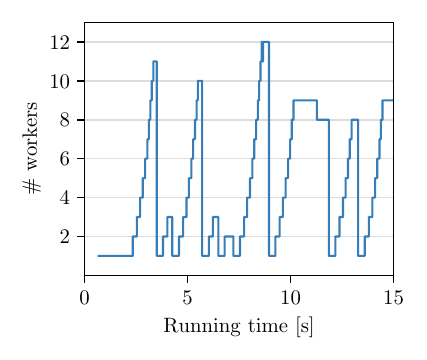}%
	\includegraphics[scale=\pyplotscale]{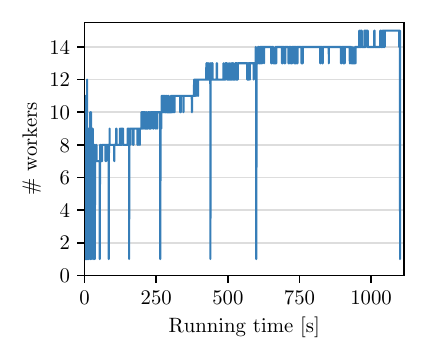}%
	\caption{Job volume (\# workers) for a difficult \texttt{Assembly} instance, in the first few seconds (left) and over the job's entire processing time (right).}
	\label{fig:mallotane-volume-history}
\end{figure}

Fig.~\ref{fig:mallotane-volume-history} illustrates how a job's volume changes over time in our experiment.
The displayed instance from the \texttt{Assembly} domain requires a total of 31 SAT solving calls and finishes within 18.4 minutes of total running time.
\lilotane{} with a sequential \cadical{} backend on this instance exceeds its time limit (30 minutes) more than 22 minutes into its 27th SAT solving call.
\mallotane{} solves that increment 263\,s after application start, taking 108\,s for the SAT solving call itself.
Whenever an increment is solved, the job shrinks down to a single ``standby'' worker (the job's root worker).
Whenever a new revision is introduced, the job's demand and consequently its volume increase exponentially over time until the demand has reached the current fair share of resources which the job is entitled to.
The most difficult increment to solve is the final unsatisfiable increment at a SAT solving time of 501.5 seconds, after which the very final (satisfiable) increment only takes 0.6\,s to solve.
In the latter half of the job's life time, the number of cores associated with the job mostly ranges between 39 (13 workers) and 45 (15 workers).

We can conclude from our experiments that connecting SAT-based applications to \mallob{} as a service-based backend can indeed be beneficial and can result in improved performance.
Naturally, the degree of improvement heavily depends on the amount of work the application performs apart from SAT solving.
In this sense, (SAT-based) hierarchical planning proved to be a challenging yet interesting application to explore since the problem domains are extremely diverse in their behavior and therefore highlight the strong points of our approach---good speedups for difficult solve calls and improved performance via flexible rescaling of jobs---as well as its remaining weakness, namely mild slowdowns on instances with no difficult SAT calls.

\section{Conclusion}  \label{lilotane-mallob:conclusion}

We explored the connection of our SAT-based hierarchical planning system \lilotane{} to our decentralized job processing and SAT solving platform \mallob{}.
We analyzed the requirements for such an undertaking and consequently integrated a low-latency interface for incremental jobs in \mallob{} and extended \mallobsat{} to support incremental SAT solving.
Our experiments showed that a combination of parallel job processing and (modestly) parallel SAT solving can result in appealing speedups in cases where our planner emits sufficiently hard SAT increments.
We consider these results encouraging for the future use of \mallob{} as a SAT backend for other applications relying on incremental SAT solving, e.g., for electronic design automation~\cite{liu2016oracle}, model checking~\cite{buning2019using}, or multi-agent path finding~\cite{surynek2022migrating}.

In terms of future work, it may be promising to go beyond \lilotane{}'s trivial sequential makespan scheduling and instead attempt to solve multiple increments of an instance in parallel.
By generalizing the incremental job model of \mallob{} and carefully extending its application interface, such an approach may also benefit other applications of incremental SAT solving (\cf{}~\cite{wieringa2013asynchronous}).
Furthermore, we intend to explore how incremental and non-incremental SAT solvers may cooperate via careful clause exchange in order to exploit the advantages of both approaches.




\bibliography{all}

\appendix

\end{document}